\documentclass[twocolumn,showpacs,preprintnumbers,amsmath,amssymb]{revtex4}
\usepackage{graphicx}% Include figure files
\usepackage{dcolumn}% Align table columns on decimal point
\usepackage{bm}% bold math

\begin{document}

%\preprint{}

\title{Normal-state Hall Angle and Magnetoresistance in quasi-2D Heavy Fermion CeCoIn$_5$ near a Quantum Critical Point }% Force line breaks with \\

\author{Y.~Nakajima$^1$, K.~Izawa$^1$, Y.~Matsuda$^1$, S.~Uji$^2$, T.~Terashima$^2$, H.~Shishido$^3$, R.~Settai$^3$, and Y.~Onuki$^3$, and H.~Kontani$^4$}
\affiliation{$^1$Institute for Solid State Physics, University of Tokyo, Kashiwanoha, Kashiwa, Chiba 277-8581, Japan}%
\affiliation{$^2$National Institute of Material Science, Sakura, Tsukuba, Ibaraki 305-0003, Japan}%
\affiliation{$^3$Graduate School of Science, Osaka University, Toyonaka, Osaka, 560-0043 Japan}%
\affiliation{$^4$Department of Physics, Saitama University,  Shimo-Okubo, Saitama 338-8570,  Japan}%

\date{\today}

\begin{abstract}

The normal-state Hall effect and magnetoresisitance (MR)  have been measured in the quasi-2D heavy fermion superconductor CeCoIn$_5$.  In the non-Fermi liquid region where the reistivity $\rho_{xx}$ exhibits an almost perfect $T$-linear dependence, the Hall angle varies as $\cot \theta_{\rm H} \propto T^2$ and the MR displays a strong violation of  Kohler's rule.  We demonstrate a novel relation between the MR and the Hall conductivity, $\Delta \rho_{xx}/\rho_{xx}\propto (\sigma_{xy}\rho_{xx})^2$.  These results bear a striking resemblance to the normal-state properties of  high-$T_{\rm c}$ cuprates, indicating universal transport properties in the presence of  quasi-2D antiferromagnetic fluctuations near a quantum critical point. 

\end{abstract}


\maketitle

	 The transport and thermodynamic properties in most metals are well described by the conventional Landau Fermi liquid theory.  Within the last decade, however,  an increasing number of strongly correlated materials, including heavy fermion (HF) intermetallics, organics, and cuprates, have been found to display striking deviations from the  Fermi liquid, when they are located close to a quantum critical point (QCP) \cite{coleman}.   In each of these materials, chemical doping, pressure, or magnetic field can tune the quantum fluctuations at zero-temperature.  As a result, these systems develop a new excitation structure and display novel thermodynamic and transport properties over a broad temperature range.  It is generally believed that the abundance of low lying spin fluctuations near the QCP gives rise to a serious modification to the quasiparticle masses and scattering cross section of the Fermi liquid.  In addition, some of these metals  show a  superconductivity.   It appears that in these superconductors many-body effects originating from the strong  spin fluctuations associated with the QCP often gives rise to unconventional superconductivity, in which Cooper pairs with angular momentum greater than zero are formed\cite{sigrist}. The relation, therefore, between unconventional superconductivity and  quantum criticality emerges as an important issue in strongly correlated systems. 
	   
	Despite extensive studies on the non-Fermi-liquid behavior in the vicinity of the QCP,  many properties in the normal state remain unresolved.   In particular the detailed transport properties, {\it i.e.} how the magnetic excitations influence the transport properties, are unsettled.   For instance, in high-$T_{\rm c}$ cuprates the normal state transport properties above the pseudogap temperature are known to be quite unusual;  the reistivity shows a $T$-linear dependence in a wide $T$-range, the Hall angle $\theta_{\rm H}$ varies as  $\cot \theta_{\rm H} \propto T^2$ \cite{ong}, and the magnetoresisitance (MR) displays a strong violation of the Kohler's rule \cite{harris}.   These non-Fermi-liquid features have been controversial  because they should ultimately be related to the mechanism of the unconventional superconductivity.    Therefore, in order to obtain deep insight into such unusual electronic transport phenomena, it is crucial to clarify whether they are universal electronic properties in the vicinity of QCP or are specific to high-$T_{\rm c}$ cuprates.
	
	Recently a new class of HF compounds with chemical formula CeMIn$_5$, where M can be either Rh, Ir, and Co,  have been discovered \cite{pet}.   Among them CeCoIn$_5$ is a superconductor with the highest transition temperature ($T_{\rm c}$=2.3~K)  among all known HF superconductors.   The normal state of CeCoIn$_5$ exhibits all the hallmarks of quantum criticality.   The two key parameters of a Fermi liquid, the electronic specific heat coefficient $\gamma=C/T$ and uniform susceptibility $\chi_{\rm 0}$, increase with decreasing temperature as $\gamma \propto -\ln T$ and $\chi_{\rm 0} \propto 1/(T+\theta)$, in marked contrast with the $T$-independent Fermi liquid behavior\cite{pet,shishido}.    Moreover  an almost perfect $T$-linear resistivity\cite{Yb,rosh} is observed from $T_{\rm c}$ up to 20~K.  These non-Fermi-liquid properties have been discussed in the light of the antiferromagnetic (AF) fluctuation near a QCP.   In fact the NMR spin-lattice relaxation rate obeys $T_1^{-1}\propto T^{1/4}$, indicating that CeCoIn$_5$ is situated near an AF instability \cite{NMR}.   Moreover the microscopic coexistence of superconductivity and static AF order has been reported in Ce(Co$_{1-x}$Rh$_x$)In$_5$ and Ce(Ir$_{1-x}$Rh$_x$)In$_5$ systems \cite{AFS}.  It should be noted that CeCoIn$_5$ has some resemblence with high $T_{\rm c}$ cuprates.  First, the electronic structure is quasi-2D, as revealed e.g. by de Haas-van Alphen  measurements\cite{shishido}.  Second, the superconducting gap symmetry most likely belongs to the $d$-wave class, indicating that the AF fluctuation plays an importnat role for the occurence of superconductivity \cite{NMR,mov,izawa}.  Third, a possible existence of the pseudogap was suggested \cite{sid}.    In general, extraction of non-Fermi liquid properties near the QCP requires high quality single crystals, since the physical properties are seriously affected by small amounts of disorder induced by the chemical doping or pressure.   CeCoIn$_5$ is suitable for such a purpose because non-Fermi liquid effect can be observed in the undoped compound at ambient pressure.    Thus CeCoIn$_5$ provides an unique opportunity for investigating the transport properties  in the presence of strong 2D AF fluctuation in the proximity of  the QCP.  
	
	In this Letter, we used the Hall effect and the MR  to clarify the microscopic scattering mechanism responsible for the unusual physical properties in the  quasi-2D HF superconductor CeCoIn$_5$.   The zero field resistivity, Hall effect and the MR, all demonstrate a spectacular breakdown of Fermi-liquid behavior and show a striking similarity to high-$T_{\rm c}$ cuprates.   These results enable us, as mentioned above, to gain strong insights into the transport properties in the presence of the quasi-2D AF  fluctuations associated with the QCP. 

	The high quality single crystals of CeCoIn$_{5}$ ($T_{\rm c}$=2.3~K) were grown by the self-flux method.   On cooling from room temperature, $\rho_{xx}(H=0)$ shows a slight increase below $\sim$200~K, followed by a crossover to metallic $T$-dependence below  $\sim40~K$ which seems to correspond to the coherent temperature $T_{\rm coh}$.   Below $T^*\simeq20$~K $\rho_{xx}$ displays an almost perfect linear $T$-dependence down to $T_{\rm c}$, as shown in the main panel of Fig.~1.   The in-plane diagonal ($\rho_{xx}$) and Hall ($\rho_{xy}$) resistivities  were measured with current {\boldmath $j$} $\parallel a$ and  {\boldmath $H$} $\parallel c$, up to 25~T and up to 14~T, respectively.  We obtained $\rho_{xy}$ from the transverse resistance by subtracting the positive and negative magnetic field data.  We have measured three different crystals and obtained similar results.
	
\begin{figure}[b]
\begin{center}
\includegraphics[width=3in]{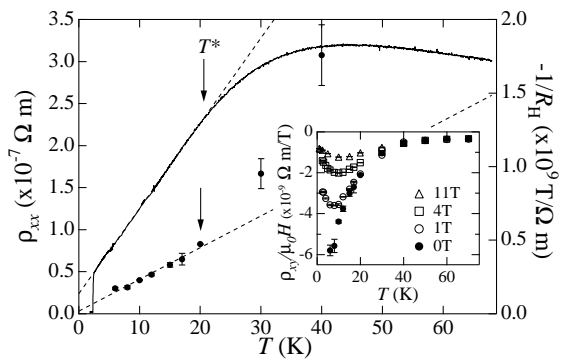}
\caption{Main panel: Temperature dependence of $\rho_{xx}(0)$ and  $1/R_{\rm H}$ in zero field limit.  The dashed straight lines are guide for eyes.  The arrows indicate $T^*\simeq 20$~K, below which $T$-linear $\rho_{xx}$ and $1/R_{\rm H}$ are observed.  Inset: $\rho_{xy}/H$ vs. $T$ at low fields. }
%\label{label}
\end{center}
\end{figure}

	We first discuss the Hall effect.  Figure 2(b) depicts the $H$-dependence of $\rho_{xy}$.    The Hall sign is negative in the present $T$- and $H$-range.   At high temperatures above 40~K,  $\rho_{xy}$ is nearly proportional to $H$, while at lower temperatures $\rho_{xy}$ shows a distinct deviation from the $H$-linear dependence.   This also can be seen by the inset of Fig.~1, in which $\rho_{xy}/H$ is plotted as a function of $T$.  The data $\rho_{xy}/H$ at $H$=0 are obtained by taking the low field limit of  $\frac{d\rho_{xy}}{dH}$.   At high temperatures, $\rho_{xy}/H$ is nearly $T$-independent.   As the temperature is lowered, $\rho_{xy}/H$ at $H$=0 decreases monotonically down to $T_{\rm c}$, while $\rho_{xy}/H$ at $H\agt$1~T turns to increase after going through a minimum above $T_{\rm c}$.  In the main panel of Fig.~1, $1/R_{\rm H}$ is plotted as a function of $T$, where $R_{\rm H}( \equiv \lim_{H\rightarrow0}\frac{d\rho_{xy}}{dH})$ is the Hall coefficient.    A $T$-linear dependence of $1/R_{\rm H}$ is observed above $T_{\rm c}$ up to $T^*\simeq$20~K, as shown by the dashed line, while a striking deviation from the $T$-linear dependence is observed at higher temperature.   It should be stressed that {\it the  temperature region of $T$-linear $1/R_{\rm H}$ coincides nicely with that of $T$-linear $\rho_{xx}$} (see arrow in Fig.1), indicating an initimate relation between $\rho_{xx}$ and $R_{\rm H}$.   The peculiar feature of the Hall effect is further pronounced by plotting the Hall angle  $\cot \theta_{\rm H} (\equiv \rho_{xx}/R_{\rm H} H$ at $\mu_0H$=1~T) vs $T^2$, as shown in Fig.~3 \cite{ong}.  The data fall on a straight line in the temperature range below $T^*\simeq$20~K and can be quite well fitted as,
\begin{equation}	
\cot \theta_{\rm H}=\alpha T^2+\beta. 
\end{equation}
A very small "residual "$\cot\theta_{\rm H}$ , $\beta \simeq0$,  is similar to the high-$T_{\rm c}$ cuprates with low impurity concentration(see Fig.~2 in Ref.\cite{ong}).   We will discuss the origin of this $T$-dependence later.

\begin{figure}[b]
\begin{center}
\includegraphics[width=2.7in]{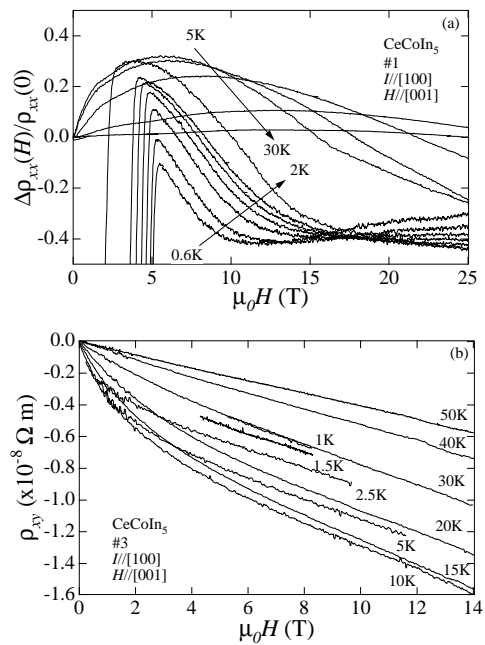}
\caption{(a)The magnetoresisitance,  $\Delta \rho_{xx}(H)/\rho_{xx}(0)$, as a function of $H$. (b) The field dependence of Hall resistivity $\rho_{xy}(H)$. }
%\label{label}
\end{center}
\end{figure}

	The Hall effect in the conventional HF systems has been studied extensively and is known to exhibit a universal behavior (see Figs.~1 and 2 in Ref. \cite{fert}).   At high temperatures exceeding $T_{\rm K}$, the Kondo temperature,  where the $f$-electrons are well localized with well-defined magnetic moments, $R_{\rm H}$ is mostly positive and is much larger than $|R_{\rm H}|$ in conventional metals.   With the development of  the Kondo coherence as the temperature is lowered, $R_{\rm H}$  decreases rapidly with decreasing $T$ after showing a broad maximum at  the temperature $T\alt T_{\rm coh}$.  At very low temperature, $|R_{\rm H}|$ becomes very small and becomes nearly $T$-independent.  These temperature dependence of $R_{\rm H}$ have been explained as follows.   Generally the Hall effect can be decomposed into two terms; $R_{\rm H}=R_{\rm H}^{\rm n}+R_{\rm H}^{\rm a}$.  Here $R_{\rm H}^{\rm n}$ is the ordinary Hall effect due to the Lorentz force and is nearly $T$-independent within the Boltzmann approximation.   Meanwhile $R_{\rm H}^{\rm a}$ represents the so-called ``anomalous Hall effect''  due to skew scattering.  The latter originates from the assymetric scattering of the conduction electrons by the angular momenta of $f$-electrons, induced by the external magnetic field.  This term is strongly temperature dependent and well scaled by the uniform susceptibility, $R_{\rm H}^{\rm a}\propto \chi_{\rm 0} \rho_{xx}$ \cite{fert,coleman2} or $\propto \chi_{\rm 0}$ \cite{onuki,upt,kon3}, above $T_{\rm coh}$.  The magnitude of $R_{\rm H}^{\rm a}$ is much larger than $|R_{\rm H}^{\rm n}|$ except at  $T \ll T_{\rm coh}$ where the contribution of the skew scattering vanishes; hence $R_{\rm H} \simeq R_{\rm H}^{\rm n}$ at very low temperatures.  

\begin{figure}[b]
\begin{center}
\includegraphics[width=2.5in]{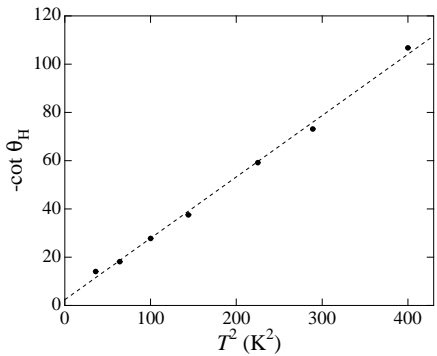}
\caption{The Hall angle $\cot\theta_{\rm H}$ as a function of $T^2$.  For details, see the text.  A fit by $\cot\theta_{\rm H}=\alpha T^2+\beta$ gives $\alpha$=0.254$(K^{-2})$ and $\beta=2.39$ (dashed line) }
%\label{label}
\end{center}
\end{figure}

	 We will show that the Hall effect in CeCoIn$_5$ shown in Figs.~1 and 2(b) is markedly different from those in ordinary HF systems.  The first important signature is that above $T_{\rm coh}\simeq$40~K, $R_{\rm H}$ is nearly $T$-independent  and can be scaled neither by $\chi_{\rm 0}\rho_{xx}$ nor by $\chi_{\rm 0}$, both of which have a strong $T$-dependence above 30~K.   In addition, although we do not show here,  $\chi_{\rm 0}$ is almost strictly $H$-linear up to 12~T at $T>$2~K, while $\rho_{xy}$ shows a nonlinear $H$-dependence below 30~K as shown in Fig.2(b).   Moreover the magnitude of  $|R_{\rm H}|$ above $T_{\rm coh}$ is much smaller than those in other HF systems at high temperatures and in the same order to $R_{\rm H}(\sim R_{\rm H} ^n)$ at very low temperaures $T \ll T_{\rm coh}$ where the skew scattering vanishes in other HF systems.  Thus both the amplitude and the $T$-dependence of $R_{H}$ are obviously in conflict with the conventional skew scattering mechanism.    These results lead us to conclude that the skew scattering in CeCoIn$_5$ is negligibly small and $T$-dependence of $R_{\rm H}$ mainly stems from the normal part of the Hall effect $R_{\rm H}^{\rm n}$.  The absence of skew scattering  has also been reported in CeMIn$_5$ \cite{sakamoto} and double layered Ce$_2$CoIn$_8$ \cite{noskew}.    At present the absence of skew scattering in these Ce compounds is an open question.   The absence of the skew scattering enable us to analyze the normal Hall effect $R_{\rm H}^{\rm n}$ in detail.   Having established the evidence of  strongly $T$-dependent $R_{\rm H}^{\rm n}$, which is incompatible with the conventional Fermi liquid theory,  we move on to  the MR.
  
\begin{figure}[b]
\begin{center}
\includegraphics[width=3in]{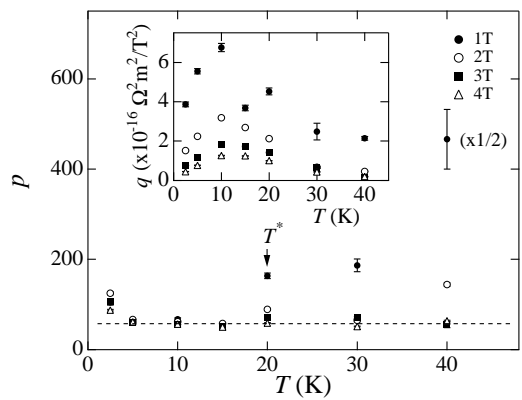}
\caption{Main panel: $p (\equiv \Delta\rho_{xx}(H)/\sigma_{xy}^2(H)\rho_{xx}^3(0))$ is plotted as a function of  $T$ at low fields.  The modified Kohler's rule represented by Eq.(5) well describes the data below $T^*\simeq$20~K.   Inset:  $q(\equiv\Delta \rho_{xx} (H) \rho_{xx}(0)/(\mu_0H)^2)$ vs. $T$ at various $H$ is plotted.  The strong violation of the Kohler's rule given by Eq.(2) is observed.  }
%\label{label}
\end{center}
\end{figure}
	  	  	  
	   The MR in most conventional HF materials also shows a similar behavior \cite{MR}.  At very low temperature $T \ll T_{\rm K}$, the MR is positive due to the orbital effect.  Then the MR decreases with temperature, changes the sign reaching the negative minimum at $T \sim T_{\rm K}/2$.   The negative MR is due to the suppression of the spin-flop scattering by magnetic field.  Figure 2 (a) depicts the MR,  $\Delta \rho_{xx}(H)/\rho_{xx}(0)$,  as a function of $H$.  A notable MR with an initial positive slope are observed at all temperaures.   At high field the MR shows a crossover from positive to negative.  The crossover field increases with temperature.  Below 2K the MR turns positive at very high field after showing a broad minimum.  Thus the MR of CeCoIn$_5$ displays a peculiar $H$-dependence, which is very different from ordinary HF systems \cite{tail}.  The positive MR at very high field below 2~K is most likely to be an usual orbital MR due to the cyclotron motion of the electrons.   The negative MR at high field seems to appear as a result of the suppression of the spin-flop scattering, similar to the other HF compounds.  The fact that the negative MR can be observed even at 25~T indicates the presence of a characteristic energy scale of the AF fluctuation which corresponds to a few meV.    We note that recent scaling argument  based on the measurements  of $C$ and $\chi_{\rm 0}$ for Ce$_{1-x}$La$_x$CoIn$_5$ suggests a presence of  the new energy scale of the excitation which is an order of $\sim$4~meV \cite{nakatsuji}.   We believe a close relation exists between the negative MR and the new eregy scale suggested in Ref.\cite{nakatsuji}.      In the conventional Fermi liquid metals the MR should obey the Kohler's rule which reads
\begin{equation}
\Delta\rho_{xx}(H)\propto\frac{H^2}{\rho_{xx}(0)}.
\end{equation}
To examine this relation, we plot $\Delta \rho_{xx} (H) \rho_{xx}(0)/H^2$ against $T$ at various $H$ in the inset of Fig.~4.    As shown in Fig.~4(a), this quantity is constant neither in $T$ nor in $H$, in strong violation of Kohler's rule. 
	  
	We are now in the  position to make a quantitative analysis for the $T$-linear resistivity, the Hall angle $\cot \theta_{\rm H} \propto T^2$ , and  the violation of the Kohler's rule in the MR, all of which signal a fundamental breakdown the Fermi-liquid behavior.    All of these peculiar transport properties reminds one of high-$T_{\rm c}$ cuprates.  In the latter materials, two opposing views have been put forth to explain these transport properties.  The first is the existence of two distinct relaxation time scales, each associated with the spinon and holon excitation in 2D CuO$_2$-planes \cite{ong,harris}.  However, the existence of such excitations would seem unlikely in the present Kondo system with different electronic structures and  different ground states,  though we cannot completely exclude this possibility.  The second is the modification of the transport quantities arising from the AF spin fluctuation near the QCP.  According to the spin fluctuation theory, the transport properties are governed by the staggered susceptibility $\chi_{\rm Q}$ \cite{kon1,kon2}.  For instance, $\rho_{xx}$ and $R_{\rm H}$ are given as, 
\begin{equation}
\rho_{xx}\sim\xi_{\rm AF}^2T^2\sim\chi_{\rm Q}T^2,
\end{equation}
\begin{equation}
R_{\rm H}\sim\pm\xi_{\rm AF}^2\sim\chi_{\rm Q},
\end{equation}
where $\xi_{\rm AF}$ is the AF correlation length \cite{kon1}.   Since $\chi_{\rm Q}$ obeys the Curie-Weiss law near the QCP, $\chi_{\rm Q} \propto 1/(T+\Theta)$, $\rho_{xx}$ is proportional to $T$ and $R_{\rm H}$ is inversely proportional to $T$, consistent with the present results.   Moreover since $\chi_{\rm Q}$ generally exhibits a non-linear $H$-dependence, in contrast to $\chi_{\rm 0}$ with a linear $H$-dependence, the  deviation from $H$-linear dependence of $\rho_{xy}$ shown in Fig.~2(b) is also consistent with Eq.(4).  The spin fluctuation theory near the AF QCP predicts that the MR is also strongly influenced by $\chi_{\rm Q}$ and Kohler's rule should accordingly be modified as \cite{kon2}, 
\begin{equation}
\frac{\Delta\rho_{xx}(H)}{\rho_{xx}(0)} \propto \left(\frac{\chi_{\rm Q}H}{\rho_{xx}(0)}\right)^2 \propto ( {\rho_{xx}(0)\sigma_{xy}(H)})^2
\end{equation}
where $\sigma_{xy}(\equiv \rho_{xy}/(\rho_{xx}^2+\rho_{xy}^2))$ is the Hall conductivity.  We now examine the MR in accordance with this ''modified Kohler's rule", which well explains the MR of high-$T_c$ cuprates \cite{harris,kon2}.   The main panel of Fig.~4 depicts $p (\equiv \Delta\rho_{xx}(H)/\sigma_{xy}^2(H)\rho_{xx}^3(0))$ vs $T$ at several fields.  Below $T^*\simeq$20~K, $p$ is nearly constant in $T$ and $H$,  providing a support for the validity of Eq.(5).     (The deviation of $p$ from the high temperature value at 2.5~K is possibly due to the influence of the superconducting fluctuation just above $T_{\rm c}$ or the pseudogap reported in Ref.\cite{sid}.)   A strong deviation can be seen above $T^*$.  Thus both the Hall effect and the MR at low field below $T^*$ in the coherent region are well described by Eqs.(4) and (5).  

	Summarizing the salient features in the normal state transport properties of quasi-2D HF superconductor CeCoIn$_5$, which is located near the QCP; (1)almost perfect $T$-linear $\rho_{xx}$, (2) $T^2$-dependent  $\cot\theta_{\rm H}$, (3)the MR which obeys a modified Kohler's rule.   All of these peculiar properties are observed  in the same temperature range  between $T_{\rm c}$ and  $T^*\simeq$ 20~K.   {\it  These are new hallmarks of the non-Fermi liquid behavior in the transport property} and well explained by the spin fluctuation theory near the QCP.   These results also bear striking resemblance to high-$T_{\rm c}$ cuprates.  It should be noted that the ground state of high-$T_{\rm c}$ cuprates is a Mott insulator while the ground state of CeCoIn$_5$  appears to be an AF metal.  The striking similarity between two different systems implies a universal excitation structure of AF fluctuation near the QCP, which promotes future study.
	
	We thank S.~Ohara, Y.~Ohkawa, I.~Sakamoto, H.~Sato, K.~Ueda, S.~Watanabe, and  V. Zlatic for valuable discussions.

\end{document}